\documentstyle[aps,epsfig,preprint]{revtex}

\begin{document}

\title{Disentangling instrumental broadening}

\author{Antonio Cervellino$^{a,b}$, Cinzia Giannini$^{b}$, 
Antonietta Guagliardi$^{b}$ and Massimo Ladisa$^{b,\star}$}
\address{
$^{a}$ Paul Scherrer Institute CH-5232, Villigen PSI, Switzerland\\
$^{b}$ Istituto di Cristallografia (IC-CNR), Via G. Amendola 122/O, I-70126, Bari, Italy}
\date{\today}
\maketitle
\begin{abstract} 
A new procedure aiming at disentangling the instrumental profile broadening 
and the relevant X-ray powder diffraction (XRPD) profile shape is presented. 
The technique consists 
of three steps: denoising by means of wavelet transforms, background 
suppression by morphological functions and deblurring by a Lucy--Richardson 
damped deconvolution algorithm. Real XRPD intensity profiles of ceria samples
are used to test the performances. Results show the robustness of the method 
and its capability of efficiently disentangling the instrumental 
broadening affecting the measurement of the intrinsic physical line profile. 
These features make the whole procedure an interesting and user-friendly 
tool for the pre-processing of XRPD data.
\end{abstract}
\noindent 
$^\star$ Corresponding author,
E\_mail: massimo.ladisa@ic.cnr.it,
Phone: +39 0805929166,
Fax: +39 0805929170

\newpage

\section{Introduction}

Precise knowledge of X-ray diffraction profile shape is crucial in 
the investigation of the microstructural properties of polycrystalline materials 
(Snyder et al., 1999). 
A reliable instrumental line-broadening analysis is thus a pre-processing step 
in most of the whole powder pattern fitting softwares. 
A new procedure aiming at disentangling the instrumental profile broadening 
and the intrinsic physical diffraction profile is presented and applied 
for the first time to XRPD data. The technique consists 
of three steps: denoising by means of wavelet transforms, background 
suppression by morphological functions and deblurring by a Lucy--Richardson 
damped deconvolution algorithm. Real XRPD intensity profiles of ceria samples
 (Balzar et al., 2004) are used to test the performances. Results show the 
robustness of the method 
and its capability of efficiently disentangling the instrumental 
broadening affecting the microstructural information. 
These features make the whole procedure an interesting and user-friendly 
tool for the pre-processing of XRPD data.

\section{The method}

Four different raw datasets were downloaded
\footnote{
www.du.edu/\%7ebalzar/s-s\_rr.htm, www.boulder.nist.gov/div853/balzar,
www.ccp14.ac.uk}
%\footnote{
%\verb|www.du.edu/\%7ebalzar/s-s\_rr.htm, 
%\verb|www.boulder.nist.gov/div853/balzar,
%\verb|www.ccp14.ac.uk}
in pairs. 
For each pair, one dataset was collected
on the annealed ceria specimen - representing the instrumental broadening -
and the other was collected on the broadened sample. The selected pairs are
those measured at the University of Birmingham (a high resolution x-ray
laboratory) and at the National Synchrotron Light Source (NSLS X3B1).

In order to extract the intrinsic  physical profile, each broadened
pattern has to be first corrected for the instrumental contribution. 
Several methods have been devised so far to deal with 
this problem. Among the others we quote the Stokes method (Stokes, 
1948),a Bayesian approach (Richardson, 1972) and the fundamental parameter 
approach (Cheary, 1992). The main drawbacks 
of these methods can be attributed to the difficulty in evaluating 
the background level, mainly due to peak overlapping. 

The technique proposed in this paper is a modified version of the 
one presented by Balzar et al, 2004. A blurred or degraded XRPD 
pattern can be approximately described by a Volterra equation 
$g = {\mathcal H} \otimes f + n$, 
where $g$ is the blurred XRD pattern and ${\mathcal H}$ is the distortion 
operator due to several causes, also called point spread function 
(PSF) in an optics/signal processing terminology; $f$ is the original XRPD 
pattern and $n$ is an additive 
noise, introduced during data acquisition, that corrupts the signal. 
Strictly speaking in a typical diffraction experiment we deal with a poissonian 
noise, that is a multiplicative noise. Moreover the Poisson distribution 
function resembles the Gauss one provided a sufficiently large statistics 
in photons counting.

Our strategy in disentangling the profile broadening out of the 
experimental sample relies on a three steps procedure which we 
sketch in the sequel.
\subsection{Denoising}
The noise was determined by means of wavelet transforms for the whole 
XRPD spectrum and subtracted prior to the background suppression.
\par\noindent
Wavelets is a well established method to perform several kinds of 
analyses on signals (see for instance Daubechies, 1992). The attention 
on this scale-based analysis from a frequency-based one (i.e. a pure 
Fourier approach) started when it became clear that an approach measuring 
average fluctuations at different scales might prove less sensitive to 
noise. Since then the wavelet domain has been growing up very quickly in 
several fields. Among them signal denoising has been deeply investigated 
and wavelets filter can be considered as the state of art on this subject.

Unlike conventional techniques, wavelet decomposition produces a family of 
hierarchically organized decompositions. The selection of a suitable level 
for the hierarchy will depend on the signal and experience. Often the level 
is chosen based on a desired low-pass cutoff frequency. At each level j, we 
build the j-level approximation ${\mathcal A}_j$, or approximation at level 
j, and a deviation signal called the j-level detail ${\mathcal D}_j$, or detail 
at level j. We can consider the original signal as the approximation at level 0, 
denoted by ${\mathcal A}_0$. 
The words approximation and detail are justified by the fact that ${\mathcal A}_1$ 
is an approximation of ${\mathcal A}_0$ taking into account the low frequencies 
of ${\mathcal A}_0$, whereas the detail ${\mathcal D}_1$ corresponds to the 
high frequency correction. One way of understanding this decomposition consists 
of using an optical comparison. 
Successive images ${\mathcal A}_1$, ${\mathcal A}_2$, ${\mathcal A}_3$ of a given 
object are built. We use the same 
type of photographic devices, but with increasingly poor resolution. The 
images are successive approximations; one detail is the discrepancy between 
two successive images. Image ${\mathcal A}_2$ is, therefore, the sum of image 
${\mathcal A}_4$ and intermediate details ${\mathcal D}_3$, ${\mathcal D}_4$, 
{\it i.e.}: 
${\mathcal A}_2 = {\mathcal A}_3 + {\mathcal D}_3 = {\mathcal A}_4 + {\mathcal D}_4 + {\mathcal D}_3$.
The organizing parameter, the scale $a$, is related to the level j by $a=2^j$. 
Since the resolution is $1/a$ then the greater the resolution, the smaller and finer 
are the details that can be reproduced. Thus the size of the details, at j$^{th}$ level, 
is proportional to the size of the region where the analysing function (wavelet) 
of the rescaled variable $x/a$ differs from zero.
\par\noindent
Generally speaking, the denoising procedure involves three steps. 
The basic version of the procedure follows the steps described below. 
\begin{itemize}
\item Decompose. Choose a wavelet, choose a level N. Compute the wavelet 
decomposition of the signal s at level N.
\item Threshold detail coefficients. For each level from 1 to N, select a 
threshold and apply soft thresholding to the detail coefficients. As to the 
thresholding, let $t$ denote the threshold. The hard threshold signal is $x$ 
if $|x| > t$, and is 0 if $|x| \leq t$. The soft threshold signal is 
$sign(x)(|x| - t)$ if $|x| > t$ and is 0 if $|x| \leq t$. As can be seen 
the hard procedure creates discontinuities at $x = \pm t$, while the soft 
procedure does not (see Birg\'e,~L. \& Massart,~P., 1997). Thus once we 
take a reference level called $J$, there are two sorts of details, those associated with 
indices  $j \leq J$ correspond to the scales $a=2^j \leq 2^J$ (the fine details) 
and the others, corresponding to $j > J$ (the coarser details). Choosing $j$ is crucial 
to define ${\mathcal A}$'s and ${\mathcal D}$'s: most of wavelet algorithms 
use an entropy-based criterion to select the most suitable decomposition of 
a given signal by quantifying the information to be gained by performing each split.
\item Reconstruct Compute wavelet reconstruction using the original 
approximation coefficients of level N and the modified detail coefficients 
of levels from 1 to N.
\end{itemize}

There are different types of wavelet families whose qualities vary 
according to several criteria. Among them we quote their support: 
the wavelets having a compact support are used in local analysis. 
Therefore in our denoising approach we used the 
Daubechies wavelets family: they are compactly supported wavelets with 
highest number of vanishing moments for a given 
support width. We address the reader to Daubechies (1992) for further 
details.
\subsection{Background suppression}
The background was determined by means of morphological transforms 
for the whole XRPD spectrum and subtracted prior to the deconvolution.
\par\noindent
Morphology is a technique of image processing based on shapes. The value 
of each pixel in the output image is based on a comparison of the 
corresponding pixel in the input image with its neighbours. By choosing 
the size and shape of the neighbourhood, you can construct a morphological 
operation that is sensitive to specific shapes in the input image. Thus 
morphological functions are used to perform common image processing tasks, 
such as contrast enhancement, noise removal, thinning, skeletonization, 
filling, and segmentation (see Serra, 1994). 
\par
In our background suppression procedure the XRPD pattern is reshaped and 
padded into a two-dimensional image by building the $m \times n$ matrix whose 
elements are taken columnwise from the one-dimensional XRPD data 
(with $m \times n \simeq N$, being $N$ the length of the XRPD data). 
Then morphological functions play 
their role on it; in particular dilation and erosion are used in combination 
to implement image processing operations. A disk with a radius of three 
pixels is used as structuring element both for erosion and for  dilation. 
As to the erosion (dilation), pixels beyond the image border are assigned 
the maximum (minimum) value afforded by the data type.The morphological 
opening removes small objects from the image while preserving the shape 
and size of larger objects in the image. The overall result is a peak 
smearing effect while the background intensity remains unalterated. 
\par
Restoring the original one-dimensional pattern provides the XRPD spectrum 
background. We compared our findings to the traditional interpolation 
method and we found a satisfactory agreement. Up to our knowledge, 
this technique has never been applied to XRPD spectrum background 
suppression and it provides a reliable and user independent 
estimate of it.
\subsection{Deblurring}
The XRPD pattern plugged in the deblurring algorithm is noise-background 
free since it has been already pre-processed by the wavelets filter + 
morphological background suppressor.
\par\noindent
As to the deblurring procedure we implement the damped Lucy--Richardson 
algorithm. This function performs multiple iterations, using optimization 
techniques and Poisson statistics. In our approach the PSF is the raw 
dataset downloaded for the ceria sample - the instrumental standard - 
resembling the instrument profile (Balzar et al., 2004).  
The algorithm maximizes the likelihood that the resulting image, 
when convolved with the PSF, is an instance of the blurred image, assuming 
Poisson noise statistics. This function can be effective when you know the 
PSF but know little about the additive noise in the image. The 
Lucy-Richardson algorithm implements several adaptations to the original 
maximum likelihood algorithm that address complex image restoration tasks. 
Using these adaptations, you can reduce the effect of noise amplification 
on image restoration, account for nonuniform image quality (e.g., bad 
pixels, flat-field variation) and improve the restored image resolution 
by subsampling. As already stressed by Balzar et al, 2004, the main 
drawbacks in applying such algorithm to the single peak deconvolution are 
the noise amplification and the peak fitting bias. Noise amplification is 
dramatically reduced by both the denoising procedure and the small (some 
five) number of iterations used in the algorithm. Moreover the damp in the 
algorithm specifies the threshold level for the deviation of the resulting 
image from the original image, below which damping occurs. For pixels that 
deviate in the vicinity of their original values, iterations are suppressed. 
As to the peak fitting bias, unlike the Balzar approach, our procedure uses 
the instrumental standard pattern (with no overlapping) as the PSF to 
deconvolve the whole XRPD pattern and then we extract the 
deconvoluted/deblurred XRPD pattern in the same range of the PSF used for 
the deconvolution itself. The rationale of this choice relies on the fact 
that while the PSF peaks have no overlap, this is not the case for the 
broadened sample peaks and, thus, the peak ranges can be defined starting 
on the annealed sample rather than the broadened one.  Moreover the discrete 
Fourier transform (DFT), used by the deblurring functions, assumes that the 
frequency pattern of an image is periodic. This assumption creates a 
high-frequency drop-off at the edges of an overlapping peaks cluster. This 
high-frequency drop-off can create an effect called boundary related ringing 
in deblurred images, that is a systematic error affecting any further 
investigation on the physical meaning of the deconvolved spectrum. To reduce 
ringing our whole pattern deconvolution, as described above, resembles an 
edgetaper function removing the high-frequency drop-off at the edge of an 
image by blurring the entire image and then replacing the center pixels of 
the blurred image with the original image. In this way, the edges of the 
image taper off to a lower frequency.

\section{Results and conclusions}

The whole procedure described above has been carried out by using few 
routines of Matlab and toolboxes implemented therein. 
In Figure 1 the results of the denoising/background suppression procedure 
described above are shown for the ceria XRPD pattern collected at Birmingham. 
Since the two different 
raw datasets for the ceria sample overlap on a reduced support, we focused 
our analysis on the first five peaks: extending the whole procedure 
described above to a more realistic XRPD pattern is straightforward. 
The dataset pair collected at NSLS X3B1, treated with the present
procedure, is analyzed within the paper submitted to Phys. Rev. B (2005).
The final result is an XRPD pattern 
with narrower peaks in the same positions of the original 
ones. The integrated intensity remains constant during the whole procedure while 
the FWHM (full width at half maximum) for each peak is significantly reduced, as 
clearly reported in Table 1. 
\par\noindent
It is worth noting that once the deblurred, noise-background free 
XRPD spectrum is convoluted back to the PSF and added to the noise+background signal singled 
out at the beginning of the procedure, the XRPD pattern resembles the original one 
with satisfactory agreement, as clearly reported on the bottom in Figure 1 (${\mathcal R}_w = 0.03$).

%\ack{Acknowledgements}

%
\begin{table}
\caption{Performance of the deblurring. Units are degrees.}
\begin{tabular}{lcccccc} 
 peak number  & \vdots & 1 & 2 & 3 & 4 & 5 \\
\hline
FWHM (before) & \vdots & 0.3281 & 0.3428 & 0.3436 & 0.3673 & 0.4056 \\
FWHM (after)  & \vdots & 0.3062 & 0.3200 & 0.3148 & 0.3433 & 0.3872 \\
\hline
\end{tabular}
\end{table}
\begin{figure}
\caption{Top: original XRPD pattern. 
Middle: XRPD pattern after denoising and background suppression.
Bottom: residue between final XRPD pattern re-convoluted to the PSF together with noise and background 
and the original XRPD pattern. On y-axes XRPD intensities 
are reported on an arbitrary units scale.}
\includegraphics*[width=0.95\textwidth]{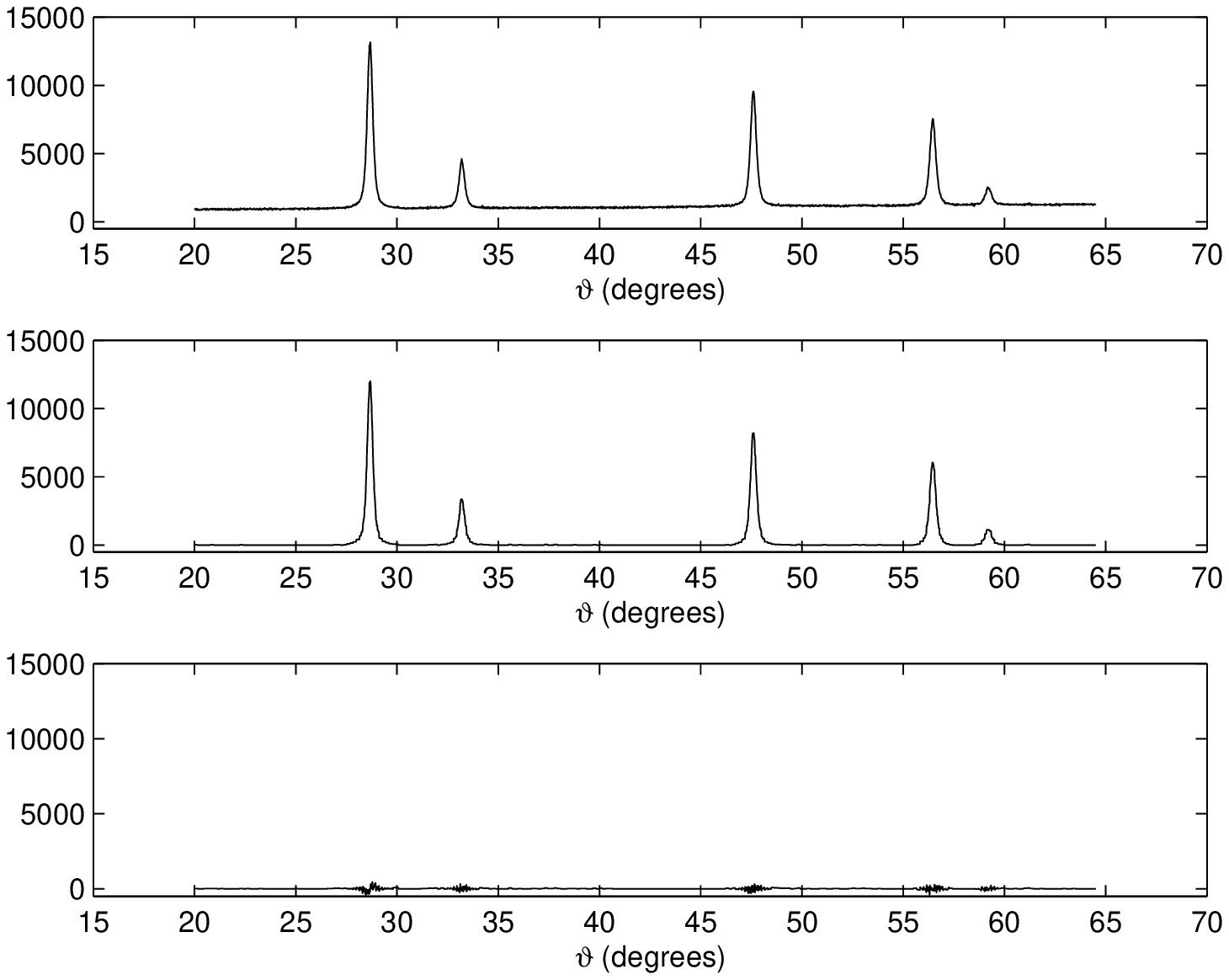}\\*
\end{figure}

\end{document}